\documentstyle{elsart}

\def \b {\beta}

\def \om {\Omega}

\def \k {{\bf k}}

\def \e {\epsilon}

\def \_t {_{\perp}}

\mathchardef\hh="7254 %calligraphic T

\begin{document}

\begin{frontmatter}
\title{The effect of small-scale forcing on large-scale structures \\
in two-dimensional flows} 
\author{Alexei Chekhlov and Steven A. Orszag}
\address{
Fluid Dynamics Research Center, Princeton University, Princeton, 
NJ 08544-0710}
\author{Semion Sukoriansky}
\address{
Department of Mechanical Engineering, Ben-Gurion University of the Negev,\\
Beer-Sheva 84105, Israel}
\author{Boris Galperin}
\address{
Department of Marine Science, University of South Florida,\\
St. Petersburg, FL 33701}
\author{Ilya Staroselsky}
\address{Cambridge Hydrodynamics, Inc., P.O. Box 1403, Princeton, NJ 08542}
\date{\today}
\maketitle

\begin{abstract}
The effect of small scale forcing on large scale structures in
$\beta$-plane two-dimensional (2D) turbulence is studied using
long-term direct numerical simulations (DNS).  We find that nonlinear
effects remain strong at all times and for all scales and establish an
inverse energy cascade that extends to the largest scales available in
the system. The large scale flow develops strong spectral anisotropy:
$k^{-5/3}$ Kolmogorov scaling holds for almost all $\phi$, $\phi =
\arctan (k_y/k_x)$, except in the small vicinity of $k_x = 0$, where
Rhines's $k^{-5}$ scaling prevails.  Due to the $k^{-5}$ scaling, the
spectral evolution of $\beta$-plane turbulence becomes extremely slow
which, perhaps, explains why this scaling law has never before been
observed in DNS.  Simulations with different values of $\beta$
indicate that the $\beta$-effect diminishes at small scales where the
flow is nearly isotropic. Thus, for simulations of $\b$-plane
turbulence forced at small scales sufficiently removed from the scales
where $\b$-effect is strong, large eddy simulation (LES) can be used.
A subgrid scale (SGS) parameterization for such LES must account for
the small scale forcing that is not explicitly resolved and correctly
accommodate two inviscid conservation laws, viz. energy and
enstrophy. This requirement gives rise to a new anisotropic stabilized
negative viscosity (SNV) SGS representation which is discussed in the
context of LES of isotropic 2D turbulence.
\end{abstract}

\end{frontmatter}

\input psfig
\section{Introduction}

Turbulent flows subjected to differential rotation develop spectral
anisotropy.  Understanding and modeling such flows present a major
theoretical and experimental challenge, mainly due to their
geophysical, astrophysical and plasma physics importance.  The
simplest two-dimensional system of this kind describes the flow of a
thin layer of homogeneous fluid on the surface of a rotating sphere.
Such flows are governed by the barotropic vorticity equation in the
$\beta$-plane approximation in which the fluid moves in tangential
planes \cite{pedlosky},
\begin{equation}
{\partial \zeta \over \partial t} + {\partial \left (\nabla^{-2} \zeta, 
\zeta \right ) \over \partial (x, y) } + \beta {\partial \over \partial x} 
\left ( \nabla^{-2} \zeta \right ) = \nu_o \nabla^2 \zeta + \xi. \label{one} 
\end{equation}
Here, $\zeta$ is the fluid vorticity, $\nu_o$ is the molecular
viscosity, and $\xi$ is the forcing; $x$ and $y$ are directed eastward
and northward, respectively.

The constant $\beta$ is the background vorticity gradient describing
the latitudinal variation of the normal component of the Coriolis
parameter, $f = f_0 + \beta y$.  Considering the energy transfer
subrange typical of classical isotropic two-dimensional turbulence
\cite{kraichnan67,kraichnan71}, we assume that the forcing $\xi$ is 
concentrated around some high wave number $k_\xi$ and is random,
zero-mean, Gaussian and white noise in time with the correlation
function 

\begin{eqnarray}
\langle \xi({\bf k},t)\,\xi({\bf k'},t') \rangle = 8\pi\,\eta\,
\delta(k^2-k_\xi^2)\, \delta({\bf k} + {\bf k'})\,\delta(t-t'),  \label{two} 
\end{eqnarray}
where $\langle \dots \rangle$ denotes ensemble average.  Thus defined,
the forcing $\xi$ supplies the system (1) with energy and enstrophy at
rates $\int \langle \vert \xi({\bf k},t) \vert^2 \rangle/k^2\,d{\bf k}
/(2 \pi)^2 =\epsilon =\eta/k_\xi^2$ and $\int \langle \vert \xi({\bf
k},t) \vert^2\rangle\, d{\bf k} /(2 \pi)^2= \eta$, respectively.  In
the inviscid limit, the corresponding total energy and enstrophy vary
in time as in the isotropic case, according to $E(t)=E(0)+\epsilon\,t$
and $\Omega(t)=\Omega(0) +\eta\,t$; the $\beta$-term does not enter
these evolution laws explicitly.

In the linear limit with no forcing, Eq. (1) describes propagation of
planetary, or Rossby waves with the dispersion relation
\begin{eqnarray}
\omega_{\bf k} = - \beta k_x / k^2,   
\end{eqnarray}
while in the nonlinear limit without the $\beta$-term, this equation
represents classical isotropic 2D turbulence. Thus, Eq. (\ref{one})
describes the interaction of 2D turbulence and Rossby waves and is
important for understanding of planetary scale geophysical processes.

The relative simplicity of Eq. (\ref{one}) has placed it at the focal
point of theoretical geophysical fluid dynamics, and much has been
learned from this model.  However, the large scale behavior of its
solutions still remains controversial; their spectral evolution laws
have not been well established, while the spectral anisotropy and the
importance of nonlinearity have received insufficient attention in the
existing literature. Here, using DNS, we address the problems of
scaling laws and spectral anisotropy of $\beta$-plane turbulence
governed by (1) and forced at small scales.  In the next section,
evolution and scaling laws of $\beta$-plane turbulence will be
discussed.  On relatively small scales the $\beta$-effect is small and
the flow is nearly isotropic.  On larger scales, the effect of the
$\b$-term becomes progressively stronger, and spectral anisotropy
develops.  Along with that, spectral evolution slows down
considerably, compared to the isotropic case.  As a result,
prohibitively long integration time is required for low wavenumber
modes to become excited even for relatively moderate resolution.  To
address this problem, LES of $\beta$-plane turbulence can be sought.
In section 3, an eddy viscosity for such LES is calculated.  This SGS
representation includes negative Laplacian term that accounts for the
unresolved small scale forcing and inverse cascade.  Thus, this SGS
representation gives rise to a stabilized negative viscosity (SNV) SGS
parameterization \cite{suk95}.  In section 4, some results of the
application of this SNV representation to isotropic 2D flows are
presented.  Finally, in section 5, we summarize our results and
conclusions.

\section{Evolution and Scaling Laws of $\beta$-Plane Turbulence}

Preliminary asymptotic analysis of Eq. (1) reveals that as $k \to
\infty$, the amplitude of the $\beta$-term decays as $k^{-1} \cos
\phi~~ [\phi = \arctan (k_y/k_x)$], so that for large $k$, the
$\beta$-effect is expected to be small and the system should behave
mostly as isotropic 2D turbulence in the energy transfer subrange
\cite{kraichnan67,kraichnan71,montgomery}. The energy spectrum
defined as $E({\bf k},t)=2\pi\,\langle \vert \zeta({\bf k},t)\vert^2 
\rangle/k$, with $\int dk\int_0^{2\pi} E({\bf k},t) d\phi/(2\pi) 
=E(t)$ being the total energy, is nearly isotropic, 
$E({\bf k},t) \approx E(k,t)$, and Kolmo-\break
\noindent gorov-like, i.e.
\begin{eqnarray}
E_K(k)=C_K\,\epsilon^{2/3}\,k^{-5/3}, 
\end{eqnarray}
where $C_K$ is the Kolmogorov constant.  On the other hand, Rossby
waves dominate as $k \to 0$. Assuming that the energy spectrum in this
limit is determined by $\beta$ and $k$ only, one finds, using
dimensional considerations \cite{rhines75}
\begin{eqnarray}
E_R(k)=C_R\,\beta^2\,k^{-5},   
\end{eqnarray}
where $C_R$, by analogy to the Kolmogorov constant, can be referred to
as the Rhines constant.  For isotropic 2D turbulence, the estimate of
the Kolmogorov constant $C_K$ from DNS and analytical theories of
turbulence is about 6.  On the other hand, the Rhines spectrum (5) has
not been observed in DNS so far so that not only no estimate of $C_R$
is available, but even the very existence of the spectrum (5) has been
in doubt.

Turbulence and waves have comparable effects when the eddy turnover
time of isotropic 2D turbulence approaches the Rossby wave period,
which occurs at a transitional wave number, $k_t(\phi)$,
\begin{eqnarray}
k_t(\phi) = k_\beta \cos^{3/5} \phi,~~k_\beta = (\beta^3 / \epsilon)^{1/5}. 
\label{five}
\end{eqnarray}
The contour (\ref{five}) in wavevector space has been termed ``the
dumb-bell shape'' by Vallis and Maltrud \cite{vallis93}; earlier, a
similar contour was called ``lazy 8'' by Holloway \cite{holloway84} in
the context of $\beta$-plane flow predictability studies.  According
to \cite{rhines75}, as $k \to k_\beta$, the inverse energy cascade
slows down, and increasing spectral anisotropy facilitates
preferential energy transfer into large scale structures with $k_x \to
0$, or zonal jets.  This conceptual picture was further developed in
later publications.  In particular, Vallis and Maltrud
\cite{vallis93} found that as $k \to k_\beta$, the dumb-bell shape
becomes a barrier for turbulence energy.  However, since the dumb-bell
shape excludes the axis $k_x = 0$, the inverse cascade can continue in
the vicinity of $k_x = 0$ facilitating the funneling of the energy
into coherent structures corresponding to zonal flows. Rhines
\cite{rhines75}, Vallis and Maltrud \cite{vallis93}, 
and Holloway \cite{holloway86} note that the spectral anisotropy
induced by the $\beta$-term can be associated with the mechanism of
generation and maintenance of zonal flows.  Zonal structures are
robust features of the simulations by Rhines \cite{rhines75}, Vallis
and Maltrud \cite{vallis93} and Panetta \cite{panetta93}, as well as
of the large scale Earth and planetary circulations.  On the other
hand, for $k < k_{\beta}$ one would expect that nonlinear transfer
becomes relatively ineffective, and the flow regimes inside the
dumb-bell shape are more prone to linear dynamics \cite{barcilon}.
Although this assumption underlines some theories, it has not been
thoroughly tested in DNS so far.  Generally, processes at $k <
k_\beta$ were avoided in the previous DNS of $\b$-plane turbulence.
Thus, not only could these processes not be properly resolved
($k_\beta$ usually did not exceed 20 or so), but the whole region $k <
k_{\beta}$ was treated rather as an area of large scale energy
dissipation due to linear (Ekman) drag or other factors.

In the present DNS, the processes at scales of the order of
$k_\beta^{-1}$ and larger are of primary interest. These DNS employ a
fully-dealiased pseudospectral Fourier method to solve Eq. (\ref{one})
in a box $\{x,y\}\in [0,2\pi] \times[0,2\pi]$ with doubly periodic
boundary conditions and zero initial data. The spatial resolution is
$512^2$ and the 2/3 dealiasing rule is used. The time discretization
is the same as in \cite{chekhlov94}. A hyperviscous term of the form
$\nu_S\,(-1)^{p+1}\,\Delta^p\zeta+\nu_L\,(-1)^{q+1}\,\Delta^{-q}\zeta$
with $p = 8$ and $q = 5$ is introduced in (\ref{one}) to ensure sharp
energy removal from both small and large wave number modes. The
forcing $\xi$ is represented by $\xi(k,t) = A_\xi\sigma_k(t)/
\sqrt{\delta t}$ for $k \in [100, 105]$ and is zero otherwise; here,
$\delta t$ is the time step, and $\sigma_k(t)$ is a discrete and
uncorrelated in time, Gaussian, random number with unit
variance. Parameter setting for DNS reported here is: $\delta t=1.0$,
$\nu_L=20.0$, $\nu_S=1.0\times 10^{-36}$, and $A_\xi=0.1$ which
results in an energy injection rate of $\epsilon
\approx 2.7 \times 10^{-13}$. Results of three DNS, with $\beta$ = 0,
$0.053$, and $0.3$, corresponding to $k_\beta$ = $0$, $56$, and $158$,
respectively, are reported here and shown in Figs. 1--5. These cases
will be referred to as Case 1, 2, and 3, respectively.  All
simulations were terminated when the energy reached the largest scales
available in the computational box and condensation started affecting
flow behavior at smaller scales \cite{smith93,smith94}.

In Figure 1, we plot the evolution of total energy and enstrophy
normalized by the corresponding values, $E_0$ and $\Omega_0$, of the
steady state, isotropic ($\beta$ = 0) case.  A striking feature is
that increasing in $\beta$ causes increasing in computational time
necessary for the effects of condensation to appear.  For example, for
$\beta$=0.3, the effect of large scale energy saturation becomes
significant at about $t/\tau_{tu} \approx 50$, where $\tau_{tu} = 2
\pi/ \sqrt{2 E_0}$ is the large scale eddy turnover time of the steady
state in the isotropic case.  For comparison, in the corresponding
isotropic DNS, the large scale energy condensate forms after only
about $4 \tau_{tu}$. In fact, the dramatic increase in computational
time necessary for the energy to reach the lowest active modes in the
system is the major factor hampering DNS of $\beta$-plane turbulence;
the problem is exacerbated with increasing resolution.  The reason why
such a long integration time is required will be explained later.

The results plotted in Figure 1 show that in agreement with the
conservation law, $E(t)$ exhibits linear in time growth. The analysis
of mode-wise energy distribution (not shown here) indicates that
increasingly longer time is required for saturation of the lower
modes.

In Figures 2--4 we plot $\k$-dependent energy spectra averaged over
small sectors $\pm \pi/12$ around $\phi=0$ and $\phi=\pi/2$.  For Case
1, as expected, the spectrum is isotropic and Kolmogorov-like.  When
$\beta \ne 0$, spectral anisotropy develops.  As Figs. 3 and 4
indicate, this anisotropy becomes more and more pronounced with
increasing $\beta$.  Although, as mentioned earlier, spectral
evolution slows down with $\beta\ne 0$, spectral anisotropy develops
relatively fast. In Figure 3 we show that for Case 2 with $k_\xi \in
[100,105]$ and $k_\b = 56$, anisotropization begins at $t \approx
\tau_{tu}$, while for Case 3 with $k_\b = 158$, anisotropization
starts immediately.  Long term integration indicates that in the
direction $\phi=0$, or $k_y = 0$, the Kolmogorov scaling (4) develops,
while for the directions $\phi = \pm \pi/2$, a scaling similar to that
of Rhines (5), $k_y^{-5}$, prevails.  Compared to Fig. 3, Fig. 4
presents results of DNS with larger $\beta$, or larger $k_\beta$,
leading to better resolution of the region $k < k_\beta$ and a more
pronounced $k_y^{-5}$ spectrum.  In Figure 5, we compare the
compensated spectrum $E(k) \b^{-2} k^5$ along $\phi = \pm\pi/2$ at
the end of integration for Cases 2 and 3, i.e., for $t/\tau_{tu} =
16.25$ and 30.12, respectively.  An important conclusion that can be
drawn from this comparison is that along $\phi = \pm \pi/2$, the
energy spectrum scales with $\b^2$.  Combined with dimensional
considerations, this result reaffirms the validity of the scaling (5)
along $\phi = \pm \pi/2$.  More detailed analysis of the anisotropic
spectra reveals that the Kolmogorov $k^{-5/3}$ scaling holds for
almost all directions $\phi$ except for the small vicinity of $\phi =
\pm \pi/2$.  It is important to emphasize that the inverse energy
transfer does not cease for any direction and any $k \le k_\beta$ and
eventually extends to the largest scales available in the system.
Rather than being a barrier, or even a soft barrier for the inverse
energy cascade, $k_\b$ only appears to be a threshold of spectral
anisotropy.  After the energy front reaches $k_\b$, some readjustment
of $C_K$ is observed; its value decreases from 6 to about 3.  This
points to the active energy exchange between the Kolmogorov and
Rhines scaling regions facilitated by nonlinear interactions and
to intensification of the Rhines's flow regime at the expense of its
Kolmogorov counterpart.

Generally, since the $k_y^{-5}$ spectrum is much steeper than the
$k^{-5/3}$ spectrum, most of energy resides in sectors closely
adjacent to $\phi = \pm \pi/2$, so that the angular-averaged energy
spectrum also obeys Rhines's $k^{-5}$ scaling (5).  Therefore, the
present DNS confirms Rhines's arguments in \cite{rhines75}. Indeed,
this is the first time that the $k^{-5}$ spectrum is observed in
DNS. Our results also show why meaningful DNS of $\beta$-plane
turbulence requires much longer integration times than the
corresponding isotropic DNS.  Indeed, from the total energy evolution
law for both isotropic and $\b$-plane turbulence, $E(t) \propto
\epsilon t$, one estimates that the time required for the energy front
to reach wave number $k$ is $t \propto E(t)/\epsilon$. For the same
$\epsilon$, the ratio of these characteristic time scales, $t_R/t_K$,
for Rhines and Kolmogorov spectra can be estimated from (4) and (5) by
angular averaging and integrating corresponding energy spectra between
$\infty$ and $k$ giving $t_R/t_K \propto (k_\beta/k)^{10/3}$.  For
Cases 2 and 3 and $k = 15$, one finds that $t_R/t_K$ is of the order
of 80 and 2500, respectively.

Here, further discussion of the physical meaning of $k_\beta$ given by (6)
would be relevant.  Other definitions of $k_\beta$ have been used in the 
literature interchangingly with (6) (see, for instance, \cite{vallis93}).
One of those definitions is due to Rhines \cite{rhines75}, $k_\beta = 
(\beta/2U)^{1/2}$, and another one is due to Holloway and Hendershott 
\cite{holloway77}, $k_\beta = \beta/\zeta$.  Here, $U$ is a characteristic  
velocity which can be identified with the square root of the total energy 
of the system, while $\zeta$ is a characteristic vorticity that can be 
identified with the square root of the total enstrophy of the system. 
For the $k^{-5}$ spectrum, it can be shown that, within numerical coefficients, 
both Rhines's and Holloway and Hendershott's definitions provide $k_\beta = k$, 
where $k$ is the smallest wavenumber attained by the energy front.  
Furthermore, taking into account the evolution law of $\beta$-plane 
turbulence, one can show that thus defined
$k_\beta$ is time dependent, $k_\beta \sim t^{-1/4}$.  Therefore, definitions
(6) and those suggested by Rhines \cite{rhines75} and Holloway and 
Hendershott \cite{holloway77} characterize different processes in 
$\beta$-plane turbulence.  While the former provides the stationary threshold 
separating the regions of nearly isotropic and strongly anisotropic, Rossby
wave dominated turbulence, the latter pertain to the largest scales available 
in the system at any given time.

To better understand the anisotropy of the energy (or enstrophy) spectrum 
of $\b$-plane turbulence, let us consider the enstrophy transfer function 
$\hh_\om(\k,t)$ derived from the enstrophy equation 
\begin{equation}
\left[ {\partial \over \partial t} + 2 \nu_o k^2 \right] \om(\k,t) =
\hh_\om (\k,t).
\end{equation}
Here, $\om(\k,t)$ is vorticity correlation function, and $\hh_\om(\k,t)$ is 
given by
\begin{equation}
\hh_\om(\k,t)= 2 \pi k\int_{{\bf p}+{\bf q}={\bf k}}
\frac{{\bf p} \times {\bf q}}{p^2}
\langle\zeta({\bf p},t)\zeta({\bf q},t)\zeta(-{\bf k},t)\rangle
\frac{d{\bf p}~d{\bf q}}{(2\pi)^2}. 
\end{equation}

Let us introduce a cutoff wavenumber $k_c$ separating explicit
$(k < k_c)$ and implicit, or SGS $(k > k_c)$ modes and consider
enstrophy exchange between all implicit modes and a given explicit
mode $\k$ in the limit $t \to \infty$.  Following \cite{kraichnan76},
this exchange is denoted by $\hh_\om(\k \vert k_c)$ and is obtained
from (8) by extending integration only over all such triangles $({\bf
k, ~p, ~q})$ that $p$ and/or $q$ are greater than $k_c$.

In Figure 6, we plot the anisotropic energy transfer $\hh_E({\bf k}
\vert k_c)$ derived from the corresponding enstrophy transfer
$\hh_\om(\k \vert k_c)$ for Case 3 with $k_c =
50$. Obviously, $\hh_E({\bf k}\vert k_c)$ describes energy transfer
from all SGS modes with $k > k_c$ to an explicit mode with the
wavenumber ${\bf k}$, $k < k_c$. As in \cite{chekhlov94}, this energy
transfer function was calculated here directly from DNS results.
Consistently with the isotropic case \cite{chekhlov94}, $\hh_E({\bf k}
\vert k_c)$ develops a cusp at $k \to k_c$.  As $k \to 0$, strong
anisotropy prevails, and most of the energy is funneled into the
sectors adjacent to $\phi = \pm \pi/2$.  On the one hand, the modes
with small $k_x$ carry most of the energy; on the other, they allow
for weak $x$-dependency of the flow field, thus maintaining nontrivial
nonlinearity which in turn sustains anisotropic transfer.  Thus, the
modes with small $k_x$ correspond to nearly one-dimensional, zonal
structures, or jets.  Such structures are clearly seen in an
instantaneous snapshot of the vorticity field in physical space at
$t=39.6 \tau_{tu}$ for Case 3 as plotted in Fig. 7; these jets have
also been a robust feature of earlier simulations
\cite{vallis93,panetta93}.  The observation that can be made based
upon the present DNS is that the number of jets is very close to the
minimal excited wavenumber in the energy spectrum; there is no obvious
linkage between the number of jets and $k_\beta$.

The zonally-averaged velocity component $U(y,t)$ is plotted in Fig. 8
for Case 3. Initially, the jets were nearly symmetric with respect to
the reflection $y \rightarrow -y$.  By $t = 39.6 \tau_{tu}$, the zonal
jets develop strong asymmetry; as also observed in
\cite{vallis93}, eastward jets are sharp and narrow while westward
jets are smooth and wide.  It is important to know whether or not
these jets are stable with respect to perturbations with nonzero
$k_x$. Although the velocity field is time-dependent, the meridional
motions are rather slow compared to zonal flows.  In this case, the
Rayleigh--Kuo inviscid stability criterion generalized for $\beta \ne
0$ requires that the profile $U(y,t)-\beta\ y^2/2$ has no inflection
points for the flow to be linearly stable.  Examination of the second
derivative $U_{yy}(y,t)$ shown in Fig. 8 demonstrates that the
Rayleigh--Kuo criterion $U_{yy}(y,t)-\beta\ne 0$ does hold, which also
agrees with the results in \cite{vallis93}.  Furthermore, as time
advances, the profile of $U_{yy}(y,t)$ develops large negative peaks
thus ensuring linear stability at later times.  As long as the effect
of the large scale drag remains small, excitation of progressively
smaller wavenumber modes leads to a diminishing number of jets, in
agreement with other simulations \cite{panetta93}. It is interesting
to note that in models of barotropic flows on a $\b$-plane with
externally imposed mean flow $U$, the difference $\b_e = \b - U_{yy}$
may decrease to zero or even become negative.  The sign of $\b_e$ is
related to interaction between the critical layer and Rossby waves and
may point to absorption ($\b_e > 0$), perfect reflection ($\b_e = 0$),
or over-reflection ($\b_e < 0$) of Rossby waves by the critical layer
(for more details, see \cite{dick70,geisler,haynes,mci,killworth}).
In the present DNS, as well as in the previous studies by Vallis and
Maltrud \cite{vallis93}, the mean flow $U$ is generated and sustained by
small scale random forcing, while $\b_e$ is always positive.

In summary, it is useful to highlight the peculiarities caused by the
nonlinear terms in barotropic vorticity equation on a $\beta$-plane.
Although the $\beta$-term does not enter the energy and enstrophy
equations explicitly, it has a profound effect on the energy spectrum
and spectral transfer.  This effect is solely due to the nonlinearity
of Eq. (1) which facilitates interaction between Rossby waves and
vorticity modes.  As a result, the energy spectrum and transfer
develop strong anisotropy for $k < k_\b$ while the inverse energy
cascade extends to ever smaller $k$.  Furthermore, inside the
dumb-bell shape (\ref{five}), where, by simple scaling considerations,
the $\beta$-effect is expected to prevail, the energy spectrum is
determined not by $\beta$ but by the presumably irrelevant parameter
$\epsilon$.  On the other hand, in the small sectors around $\phi =
\pm \pi/2$ outside the dumb-bell shape naive considerations show that the
$\b$-term is vanishing [$\beta \partial_x \nabla^{-2} \zeta \to i
\beta k_x k^{-2} \zeta \to 0$ as $k_x \to 0$ in (1)] and the mechanism of 
anisotropic inverse transfer funnels energy into zonal jets. In the
present DNS we find that the zonal jets do indeed form, but the effect
of the $\beta$-term does not diminish by any means: the spectrum of
energy in the regions $\phi \to \pm \pi/2$ is determined by $\beta$.

The possibility of energy transfer from Rossby waves to zonal flows has
long been discussed in the literature and by no means is a trivial matter.
Straightforward generalization of the Fjortoft theorem \cite{lesieur90}
shows that a resonant triad of wavevectors one of which is aligned with 
the axis $y$ does not allow for direct energy transfer into zonal flows.  
However, the vector with $k_x = 0$ facilitates energy exchange between the
other two vectors - members of the triad.  This fact was proved in 
\cite{gill67} in much more general way; it was shown that the coupling 
coefficient is zero for any such triad that $k_x = 0$ for one of the vectors.  
Later, Newell \cite{newell69} showed that Rossby waves can transfer energy into 
zonal flows through a sideband resonance mechanism or through a quartet 
resonance.  Reinforcing this result, the present DNS indicate that not only 
energy transfer into zonal flows due to nonlinear interaction is possible, 
but in fact it is the dominant mechanism of $\beta$-plane turbulence on large
scales.

\section{Parameterization of Eddy Viscosity for LES of $\b$-Plane
Turbulence}

Since DNS of $\beta$-plane turbulence in the energy transfer subrange
demands excessive computational resources even for relatively moderate
resolution, LES could be a viable alternative.  For such LES, however,
a SGS representation is crucial since not only should it accommodate
inverse energy transfer but also it should account for the small scale
energy forcing excluded in the explicit modes.  Following
\cite{suk95}, an SGS representation for high wavenumber forced
$\b$-plane turbulence can be given by an anisotropic two-parametric
viscosity, $\nu (\k \vert k_c)$, defined by
\begin{equation}
\nu (\k \vert k_c) = - {\hh_\om (\k \vert k_c) \over {2 k^2 \om(\k)}}. 
\end{equation}
This two-parametric viscosity was calculated analytically using the
renormalization group (RG) theory of turbulence \cite{suk94}; it is
shown in Fig. 9 for the set of parameters corresponding to Case 2 of
DNS. The behavior of the theoretically derived $\nu (\k \vert k_c)$ is quite 
similar to that obtained from DNS as shown in Fig. 10 also for Case 2.  For 
large $k$, the $\b$-effect is weak, and $\nu(\k \vert k_c)$ behaves
similarly to the isotropic case \cite{suk94,chekhlov94}; there is a
sharp positive cusp and then $\nu(\k \vert k_c)$ becomes negative.  As
$k \to 0$, the effect of the $\b$-term becomes stronger; $\nu(\k \vert
k_c)$ remains negative in the vicinity of $\phi = \pm \pi/2$ but
increases to zero in other directions. The negativity of $\nu(\k \vert
k_c)$ along $\phi =\pm \pi/2$ is consistent with the strong energy
flux into zonal jets as discussed in the previous section.  It is important
to mention that when the DNS integration was terminated, the low wavenumber
end of the spectrum was not yet fully energy saturated.  This fact explains
the sharp increase in DNS-derived $\nu (\k \vert k_c)$ at small ${\bf k}$ 
evident in Fig. 10.  The sign-changing structure of $\nu (\k \vert k_c)$ is 
indicative of a physical space SGS representation which combines a negative 
Laplacian and a positive (dissipative) hyperviscosity thus presenting an
anisotropic version of the Kuramoto--Sivashinsky equation or SNV SGS
representation \cite{suk95}. The main ideas of the SNV formulation are
discussed in the next section as applied to LES of more simple
isotropic 2D turbulence.  The isotropic SNV SGS representation can be
useful for LES of $\b$-plane turbulence with $k_\b < k_c$ since
deviation from isotropy is expected only for $k < k_\b$ where, due to
the factor of $k^2$ on the right hand side of (10) below, the SGS
contribution is small.

\section{SNV SGS Representation for Isotropic 2D Turbulence}

Specific difficulties of LES of 2D turbulence in the energy transfer
subrange are that the SGS representation must simultaneously be
consistent with the conservation laws for both energy and enstrophy
(or potential enstrophy) \cite{suk95}. In addition, in LES of the
inverse energy cascade, the wavenumber modes of the energy source lie
in the subgrid scales so that the SGS representation must assume the
function of the forcing. The SGS schemes being used to date in both 3D
and quasi-2D flows usually employ Laplacian or higher order
hyperviscosities that ensure SGS energy dissipation but not forcing.
While such SGS are consistent with the dynamics of 3D turbulence where
energy is directly cascaded from large to small scales where it
dissipates, they cannot be expected to work well in quasi-2D flows
where energy cascades from small to large scales while the enstrophy
is transferred in the opposite direction.  A possible resolution to
this problem can involve the replacement of the SGS forcing by force
located in the explicitly resolved region near $k_c$.  However, this
solution is not only quite cumbersome but also significantly distorts
the explicit scales near $k_c$. In addition, this approach is
difficult for implementation in physical space, particularly for
bounded systems and/or systems with spatially nonuniform energy
sources.

An alternative approach is offered by the SNV representation developed
in \cite{suk95} for isotropic 2D turbulence.  In this approach,
Eq. (1) (with $\b$ = 0) that includes small scale forcing $\xi$ is
replaced by LES equation
\begin{eqnarray}
{\partial \zeta ({\bf k}) \over \partial t} &&+ \int_{|p|,|k-p| < {k_c}}
{{\bf p} \times {\bf k} \over {p^2}}
\zeta({\bf p})\zeta({\bf k} - {\bf p})
{d{\bf p}\over {(2\pi)^2}}
=  -\nu(k \vert k_c) k^2 \zeta({\bf k}), \\
&&0 < k < k_c.  \nonumber
\end{eqnarray}

Here, eddy viscosity is identified with the two-parametric viscosity
$\nu( k \vert k_c)$ defined by (9) (where, due to isotropy of the
flow, the angular dependence is omitted). In \cite{suk95} it was
found that this eddy viscosity is given by
\begin{equation}
\nu( k \vert k_c) = {\e \over 0.8 \Omega(t)} N(k/k_c),
\end{equation}
where $\Omega(t)$ is the total enstrophy of the system and $N(k/k_c)
\equiv \nu( k \vert k_c) / \vert \nu(0\vert k_c) \vert$ is plotted in
Fig. 11. This formulation is designed to provide constant energy
transfer with the rate $\e$. Consistent with the eddy viscosity approach,
the SGS representation (11) is a function of the flow. In a
statistical steady state, Eq. (11) coincides with the two-parametric
viscosity derived from the RG theory \cite{suk95}.

In a series of LES that employ the formulation (11) together with a
specially designed large scale drag that effectively removes energy
that reaches the largest computational scales, it was found that
simulations can be carried out nearly indefinitely (being terminated
after about 200 turnover times), while a very robust Kolmogorov
$k^{-5/3}$ energy spectrum was established.

The formulation (11) can be simplified to make it more easily
adaptable for simulations of quasi-2D flows in physical space. First,
a two-term power series expansion in powers of $k/k_c$ that satisfies
all necessary conservation laws can be derived from (11) \cite{suk95},
\begin{equation}
\nu( k \vert k_c) = {25 \over 18} {\e \over \Omega(t)} \left[ -1 +
{8 \over 5} \left({k \over k_c} \right)^2 \right].
\end{equation}
Here, the negative term in the square brackets accounts for the effect
of the small scale forcing and inverse energy cascade. The second term
describes dissipative processes and is equivalent to a biharmonic
viscosity in physical space. Then, (12) can be further simplified if
$\om$ in its dissipative term is replaced by the value obtained from
the Kolmogorov spectrum (4). Finally, the resulting SGS representation
for 2D turbulence in the energy transfer subrange takes the form
\begin{equation}
\nu( k \vert k_c) = - {25 \over 18} {\e \over \Omega(t)} + 0.511 \e^{1/3} 
k_c^{-10/3} k^2.
\end{equation}
In Figs 12 and 13, we plot the results of LES of 2D turbulence with
the SGS representation (13). One can see that the total energy and
enstrophy oscillate slightly around their steady state values while a
robust Kolmogorov energy spectrum is established.

The corresponding physical space representation of the SGS operator for
quasi-2D flows in the energy subrange is obtained using the inverse
Fourier transform of (13),
\begin{equation}
- {\partial \over \partial x_i} \left ( A_2  {\partial \over \partial x_i}
\right ) - A_4 {\partial^4 \over \partial x_i^2 \partial x_j^2 },
\end{equation}
where
\begin{eqnarray}
A_2 &&= {25 \over 18}~ {\e \over \om({\bf x})},  \\ 
A_4 &&= 0.511 \e^{1/3} (\Delta /2 \pi)^{10/3}, 
\end{eqnarray}
and where $\om({\bf x})$ denotes the enstrophy averaged over an area
adjacent to the grid cell, $\Delta$ is the grid resolution, the
Laplacian term in (14) is written in the conservative form.  Since
(14) includes two terms, a {\it negative} Laplacian and positive (in
the sense of dissipation) biharmonic, it is referred to as a {\it
stabilized negative viscosity} (SNV) formulation.  Generally, SNV
equations are far more complicated than equations of the
Kuramoto--Sivashinsky type \cite{gama91,gama94} because in the former,
the coefficients are not constant but, as in the classical eddy
viscosity approach, are functions of the flow.

\section{Conclusions}

In this paper, we describe results of direct numerical simulations of
$\b$-plane turbulence forced at small scales. As for isotropic 2D
turbulence, nonlinear interactions play a crucial role in the
evolution and dynamics of $\b$-plane turbulence on all scales and at
all times.  The effect of the small scale forcing propagates to ever
larger scales due to the inverse energy cascade.  On relatively small
scales, $k > k_\b$, the flow is nearly isotropic and behaves like
classical 2D turbulence. For modes with eddy turnover times comparable
with the Rossby wave period, the turbulence energy spectrum begins to
develop anisotropy that increases for $k \to 0$. It is found that for
almost all $\phi$ except for the small vicinity of $\phi=\pm\pi/2$,
the spectrum of turbulence is Kolmogorov-like, $E(k) \sim k^{-5/3}$.
Around $\phi=\pm\pi/2$, or $\vert k_x \vert \to 0$, the two-parametric
energy transfer function $\hh_E(\k|k_c)$ attains a sharp maximum,
which indicates that there is a strongly anisotropic energy transfer
into quasi-one-dimensional (zonal) structures.  These zonal flows are
generated and sustained by nonlinear processes, particularly,
anisotropic inverse energy cascade. For $\phi\to\pm\pi/2$, the energy
spectrum becomes very steep, $E(k) \sim k^{-5}$, as was first
suggested by Rhines \cite{rhines75}. The steepening of the spectrum
slows down flow evolution, to the extent that DNS even with relatively
moderate resolution becomes computationally prohibitive. This
difficulty can be resolved through LES of $\b$-plane turbulence. For
such LES, an SGS representation has been designed based upon a
two-parametric viscosity. We have applied this approach for an example
of isotropic 2D turbulence for which case the SGS representation is
given by the SNV parameterization.

A.C. acknowledges fruitful discussions with V. Yakhot. This work has
been supported by ARPA/ONR under Contract N00014-92-J-1796, ONR under
Contract N00014-92-J-1363, NASA under Contract NASS-32804, and
the Perlstone Center for Aeronautical Engineering Studies.  The 
computations were performed on Cray Y-MP of NAVOCEANO Supercomputer 
Center, Stennis Space Center, Mississippi.

\newpage

\begin{figure}
\centerline{\psfig{file=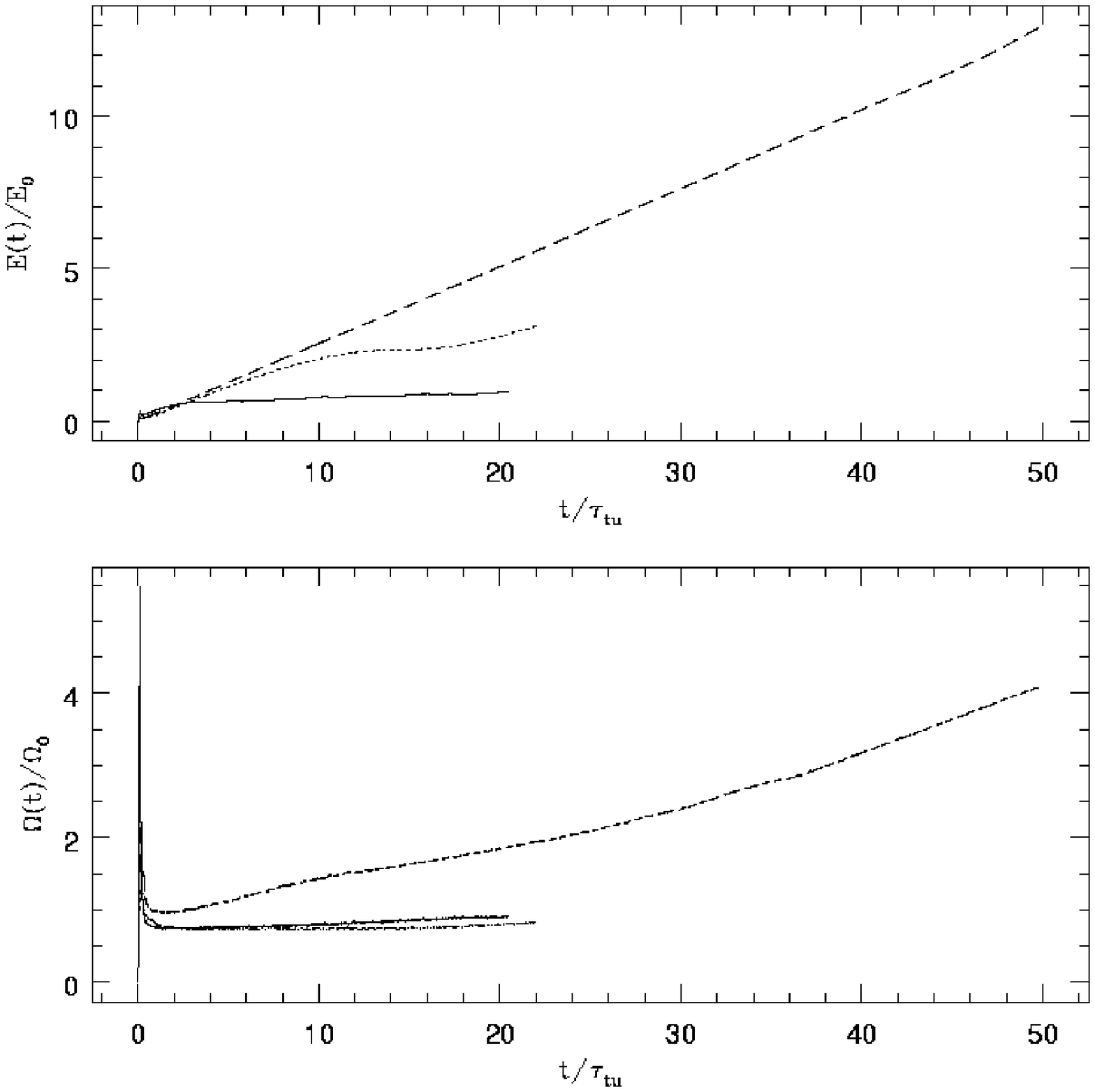,width=6.0in,height=6.0in}} 
\caption{Evolution of total energy $E(t)$ (top) and enstrophy 
$\Omega(t)$ (bottom) for Cases 1, 2 and 3 (solid, dotted, and dashed
lines, respectively).}
\label{fig1}
\end{figure}

\begin{figure}
\centerline{\psfig{file=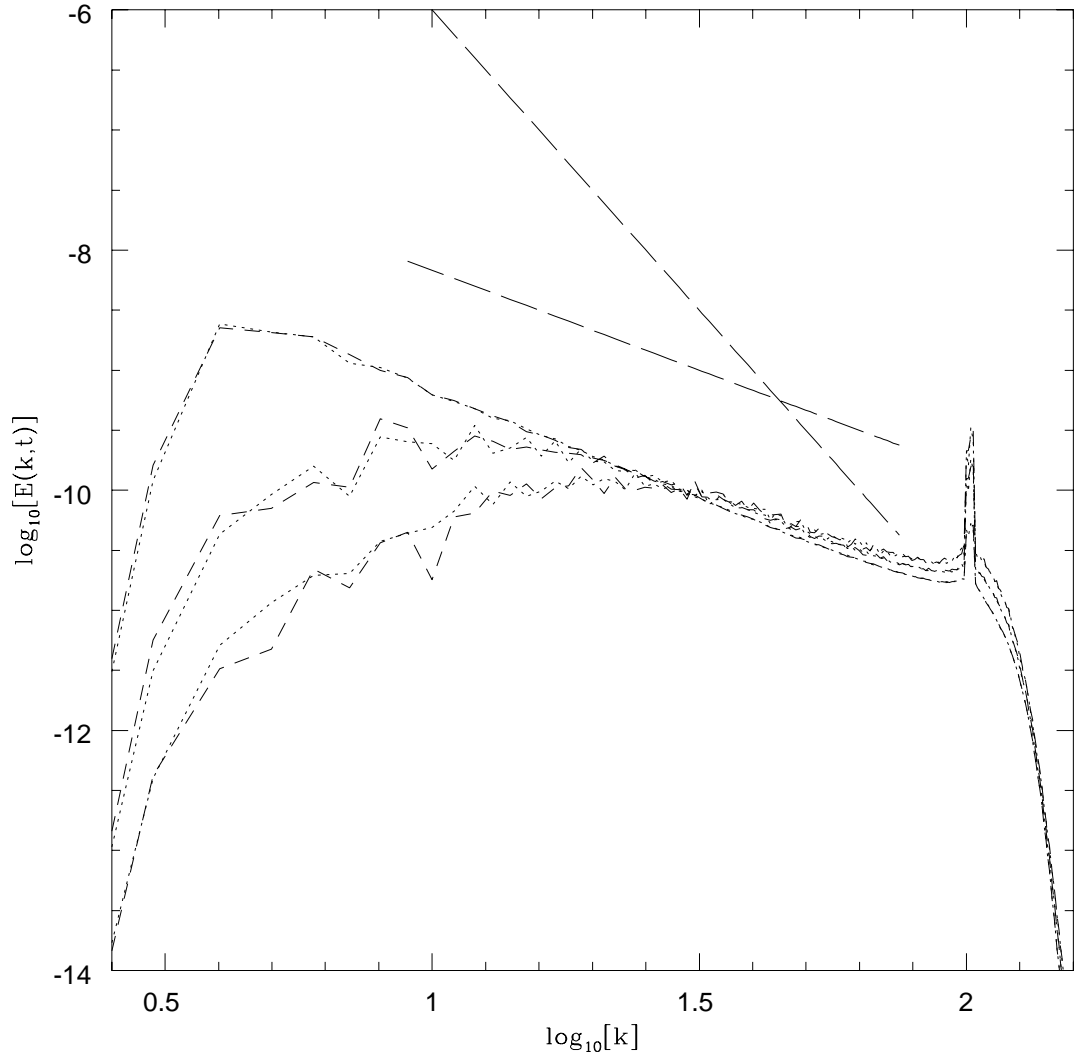,width=6.0in,height=6.0in}} 
\caption{Energy spectra for $\phi=0$ (dotted line) and $\phi=\pi/2$
(dashed line) averaged in time and over a small surrounding sector
$\pm\pi/12$ for $t/\tau_{tu}$ = 0.63, 1.19, and 8.32 for $\b$ = 0
(Case 1) (due to the central symmetry, only these two directions are
shown). Long dashed lines depict slopes of $-5$ and $-5/3$.}
\label{fig2}
\end{figure}

\begin{figure}
\centerline{\psfig{file=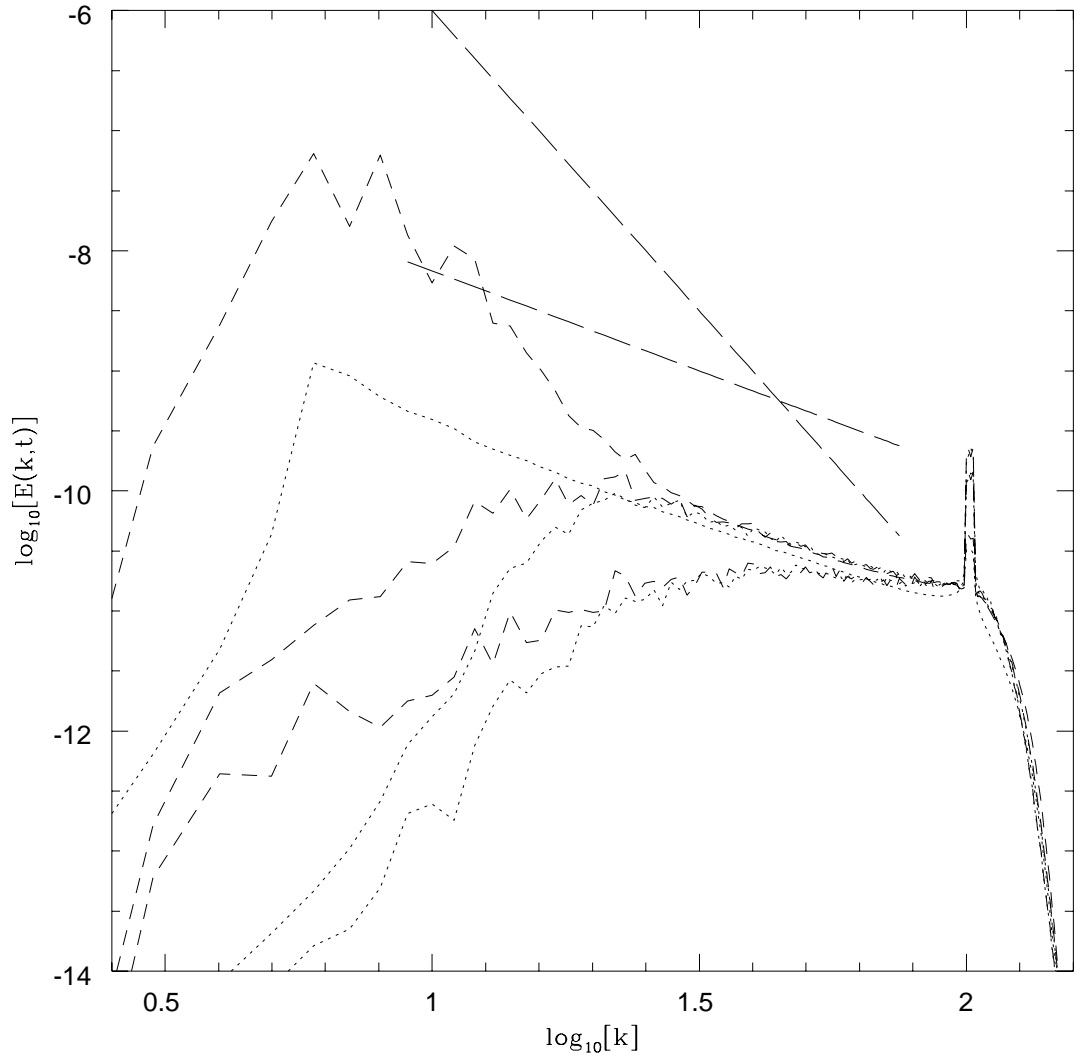,width=6.0in,height=6.0in}} 
\caption{Same as in Fig. 2 but for $t/\tau_{tu}$ = 0.63, 1.19, and
16.25 and $\b$=0.053 (Case 2).}
\label{fig3}
\end{figure}

\begin{figure}
\centerline{\psfig{file=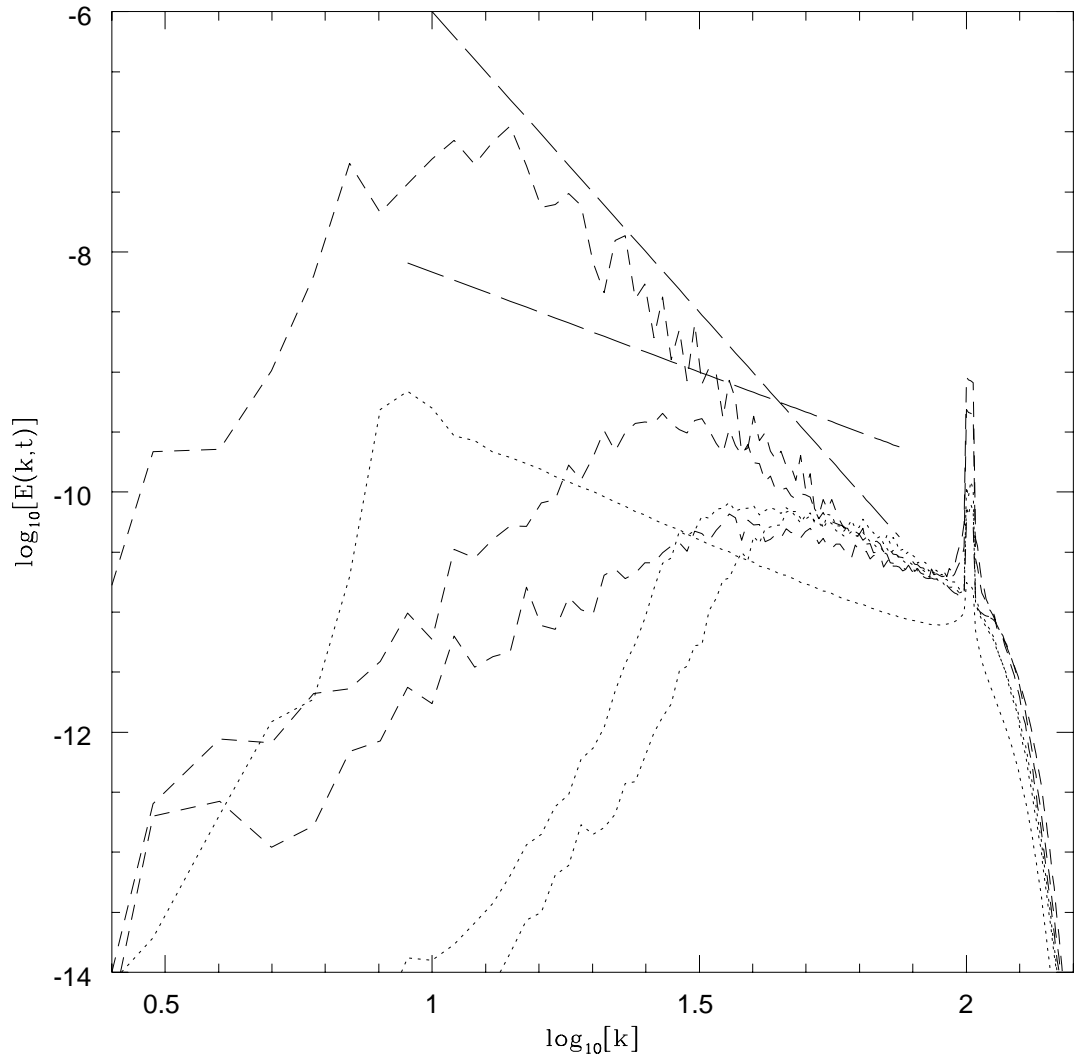,width=6.0in,height=6.0in}} 
\caption{Same as in Fig. 2 but for $t/\tau_{tu}$ = 0.63, 1.19, and
30.12 and $\b$=0.3 (Case 3).}
\label{fig4}
\end{figure}

\begin{figure}
\centerline{\psfig{file=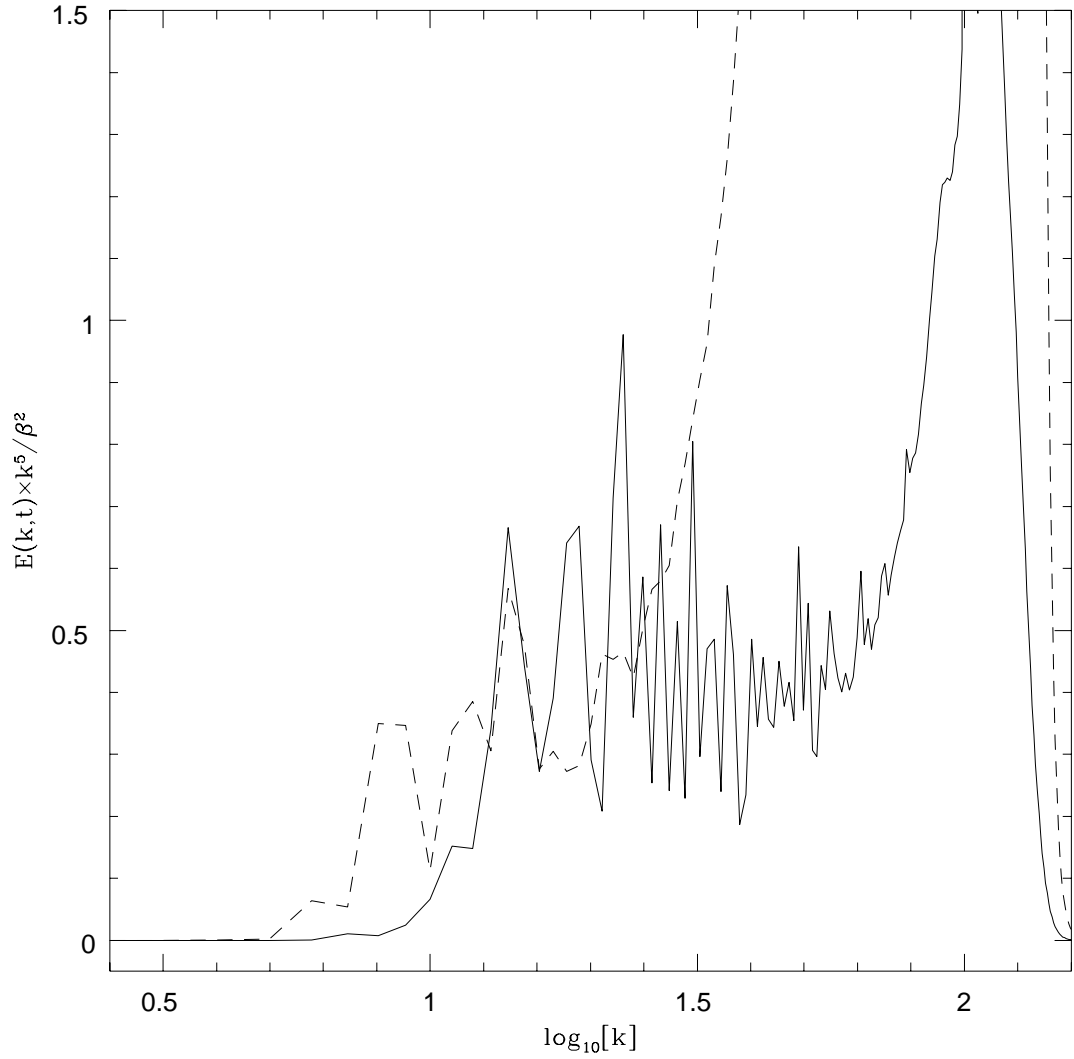,width=6.0in,height=6.0in}} 
\caption{Compensated energy spectrum for Cases 2 (dashed line) and 3
(solid line) for $t/\tau_{tu}$ = 16.25 and 30.12, respectively.}
\label{fig5}
\end{figure}

\begin{figure}
\centerline{\psfig{file=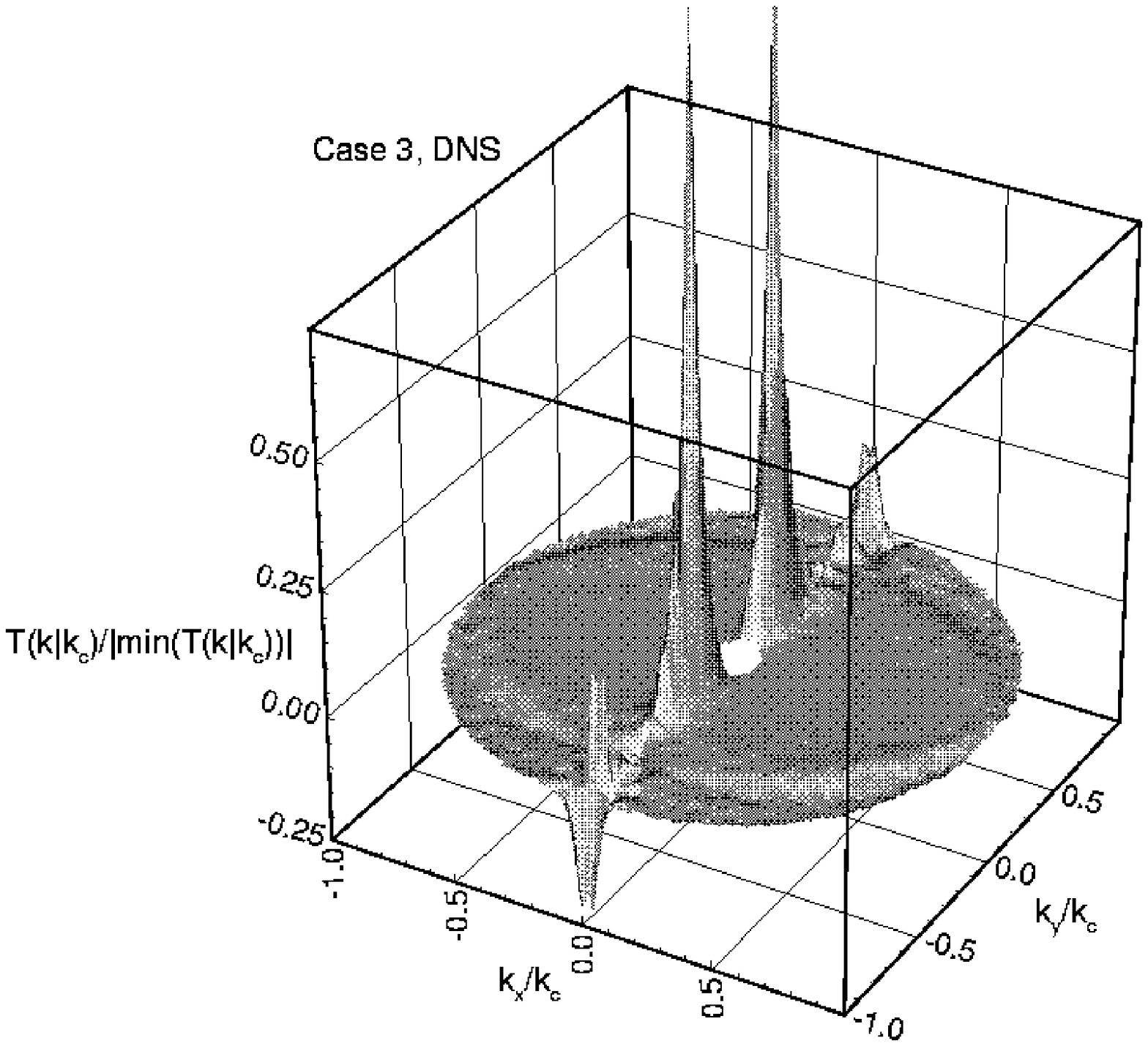,width=8.0in,height=8.0in}} 
\caption{Spectral energy transfer, $\hh_E({\bf k} \vert k_c)$, for
$k_c = 50$.}
\label{fig6}
\end{figure}

\begin{figure}
%\centerline{\psfig{file=field.ps,width=8.0in,height=8.0in}} 
\caption{Instantaneous vorticity field $\zeta({\bf x},t)$ at 
$t/\tau_{tu}=39.6$. ({\it Too big for this archive, available upon
request.})}
\label{fig7}
\end{figure}

\begin{figure}
\centerline{\psfig{file=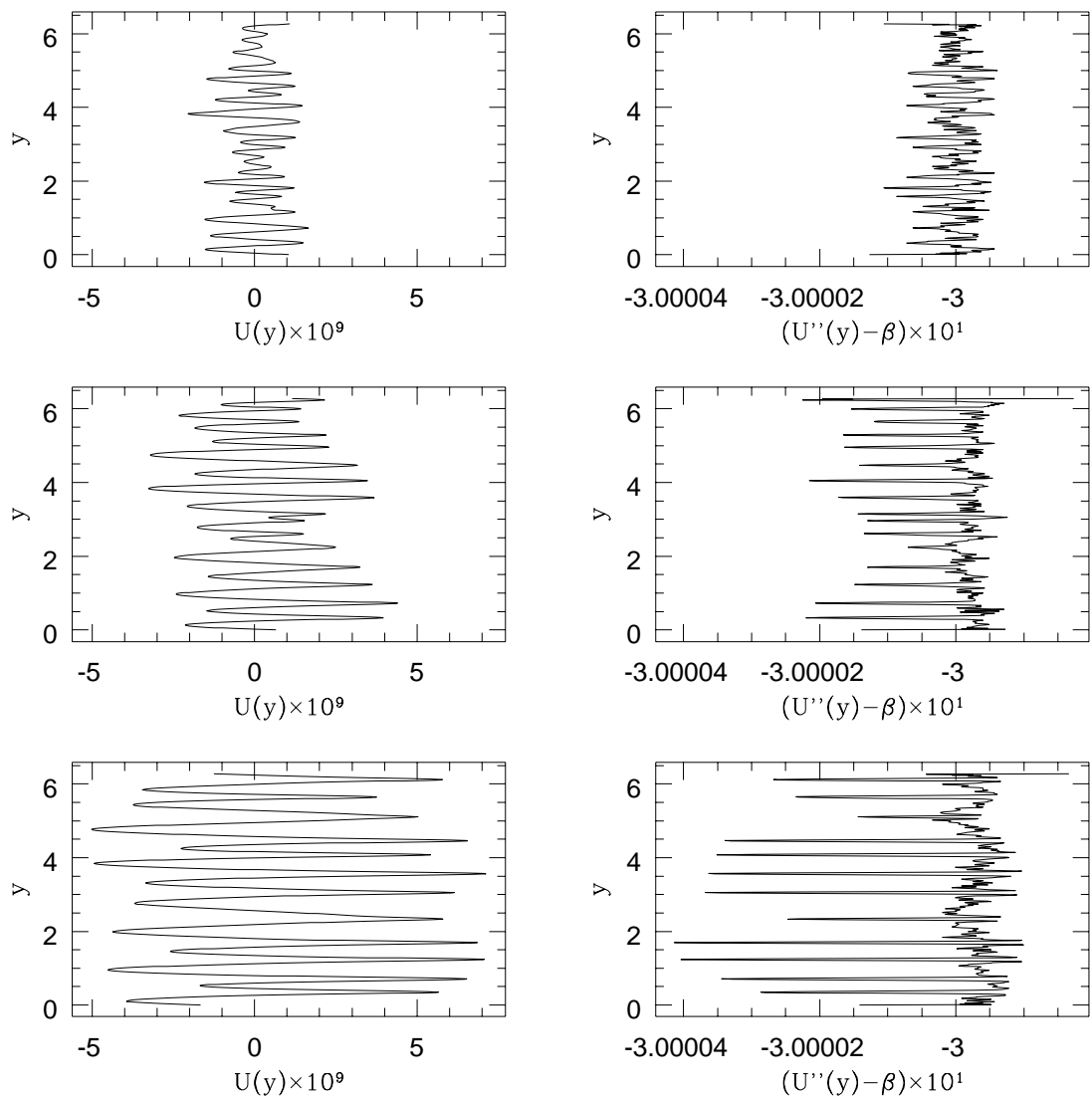,width=6.0in,height=6.0in}} 
\caption{Zonally-averaged velocity profile $U(y,t)$ (left column) 
and its second derivative $U_{yy}(y,t)$ (right column) for Case 3
at (from top to bottom) $t/\tau_{tu}$= 21.8, 39.6, and 59.4.}
\label{fig8}
\end{figure}

\begin{figure}
\centerline{\psfig{file=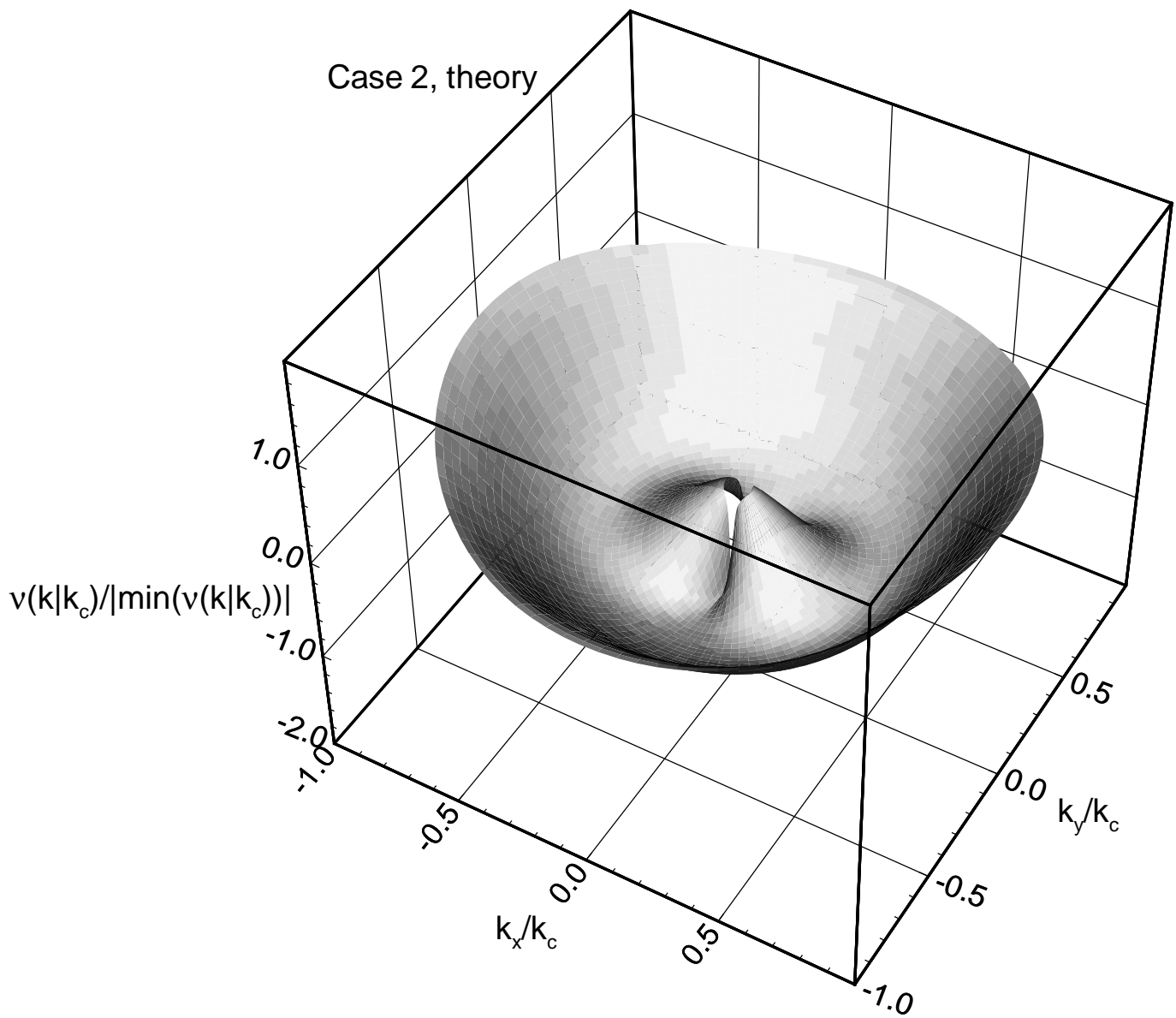,width=8.0in,height=8.0in}} 
\caption{RG-derived two-parametric viscosity normalized by the absolute
value of its minimum for $\b$-plane turbulence. Parameter setting
corresponds to Case 2 DNS.}
\label{fig9}
\end{figure}

\begin{figure}
\centerline{\psfig{file=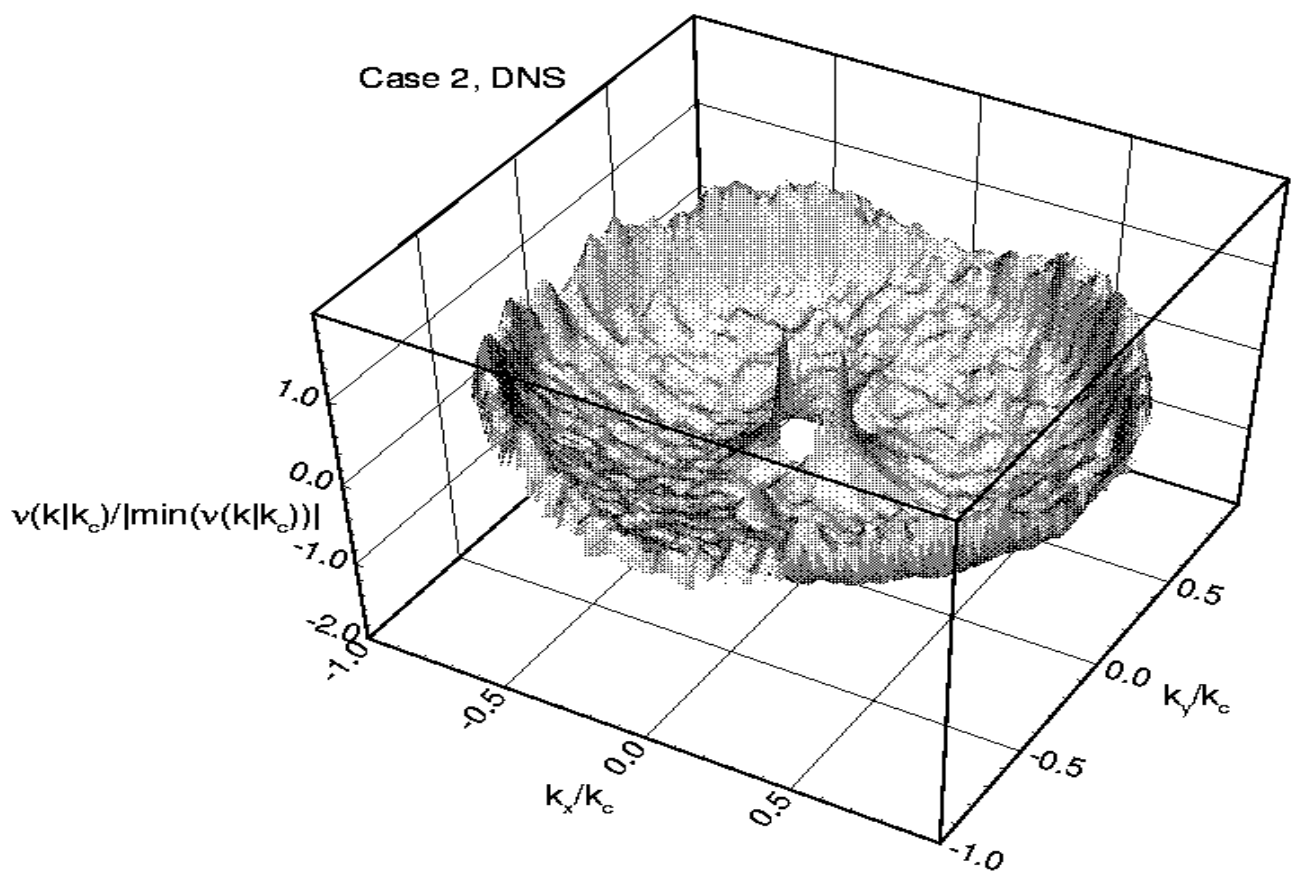,width=8.0in,height=8.0in}} 
\caption{DNS-based two-parametric viscosity normalized by the absolute
value of its minimum for Case 2 of $\b$-plane turbulence. }
\label{fig10}
\end{figure}

\begin{figure}
\centerline{\psfig{file=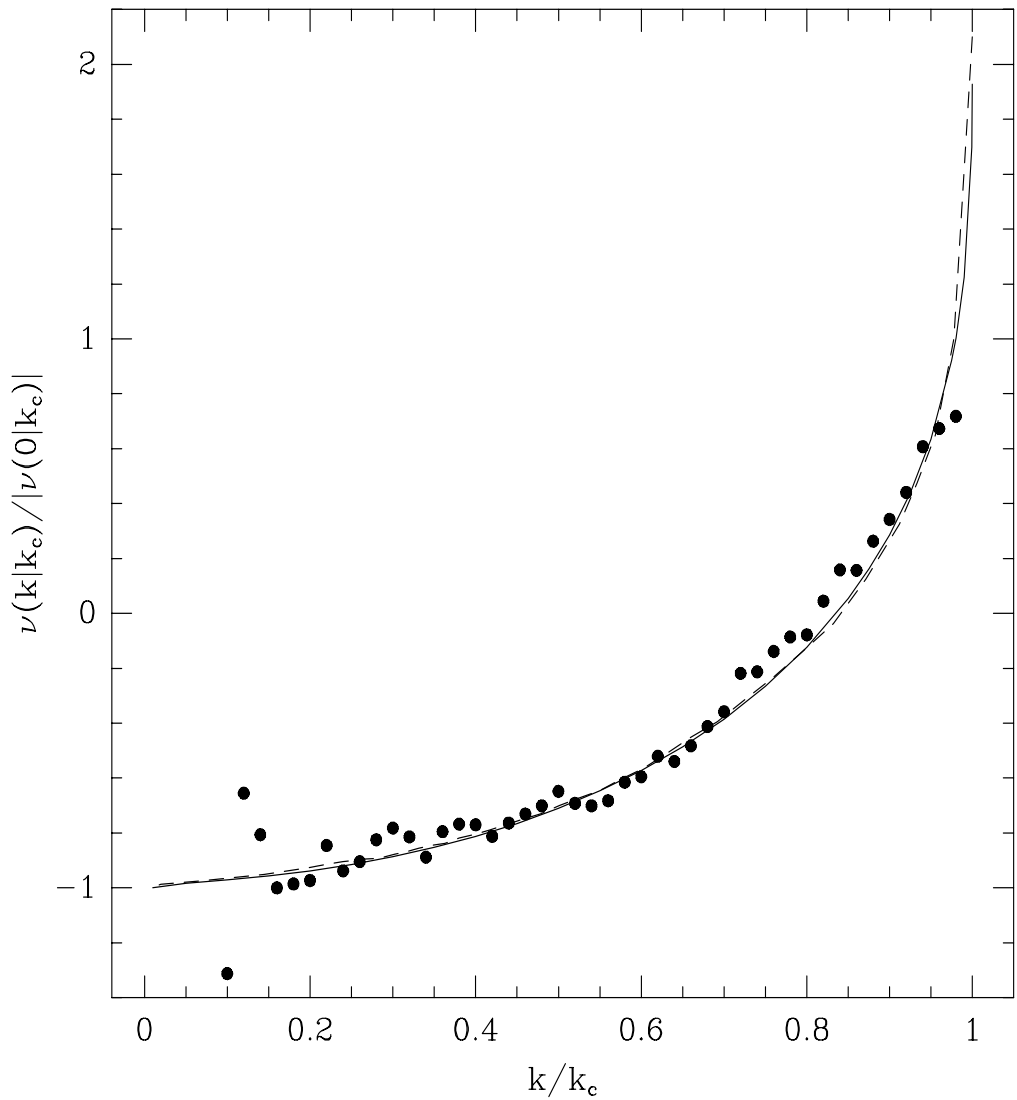,width=6.0in,height=6.0in}} 
\caption{Function $N(k/k_c)$ obtained from DNS (dots), TFM
(dashed line) and RG (solid line) (from [12]).}
\label{fig11}
\end{figure}

\begin{figure}
\centerline{\psfig{file=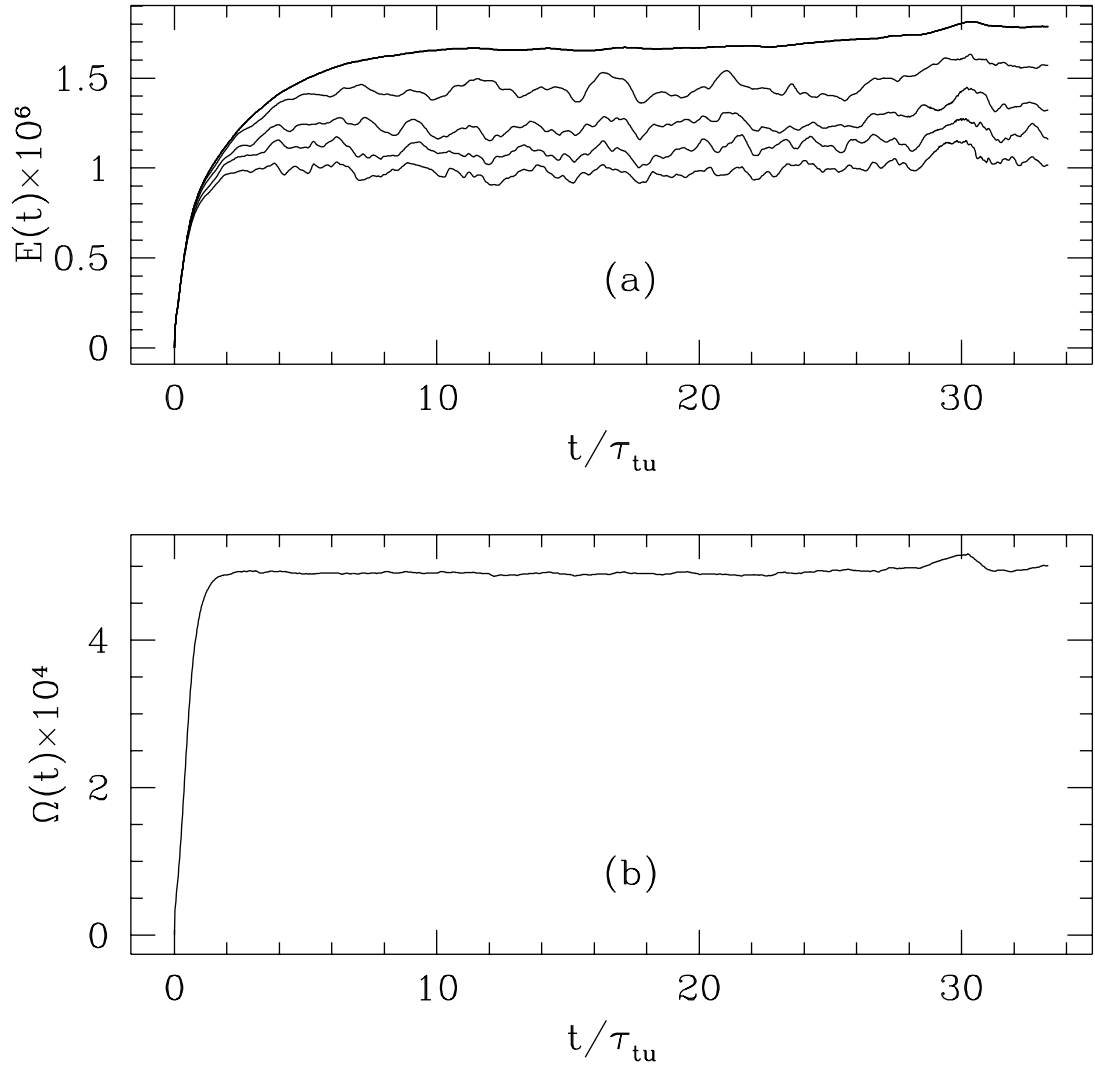,width=6.0in,height=6.0in}} 
\caption{Evolution of the total energy $E(t)$ (a) and enstrophy 
$\om(t)$ (b) in LES of isotropic 2D turbulence. In (a) we also plot
the evolution of $E(t)$ with the energy of the 4th, 5th, 6th and 7th 
modes removed.}
\label{fig12}
\end{figure}

\begin{figure}
\centerline{\psfig{file=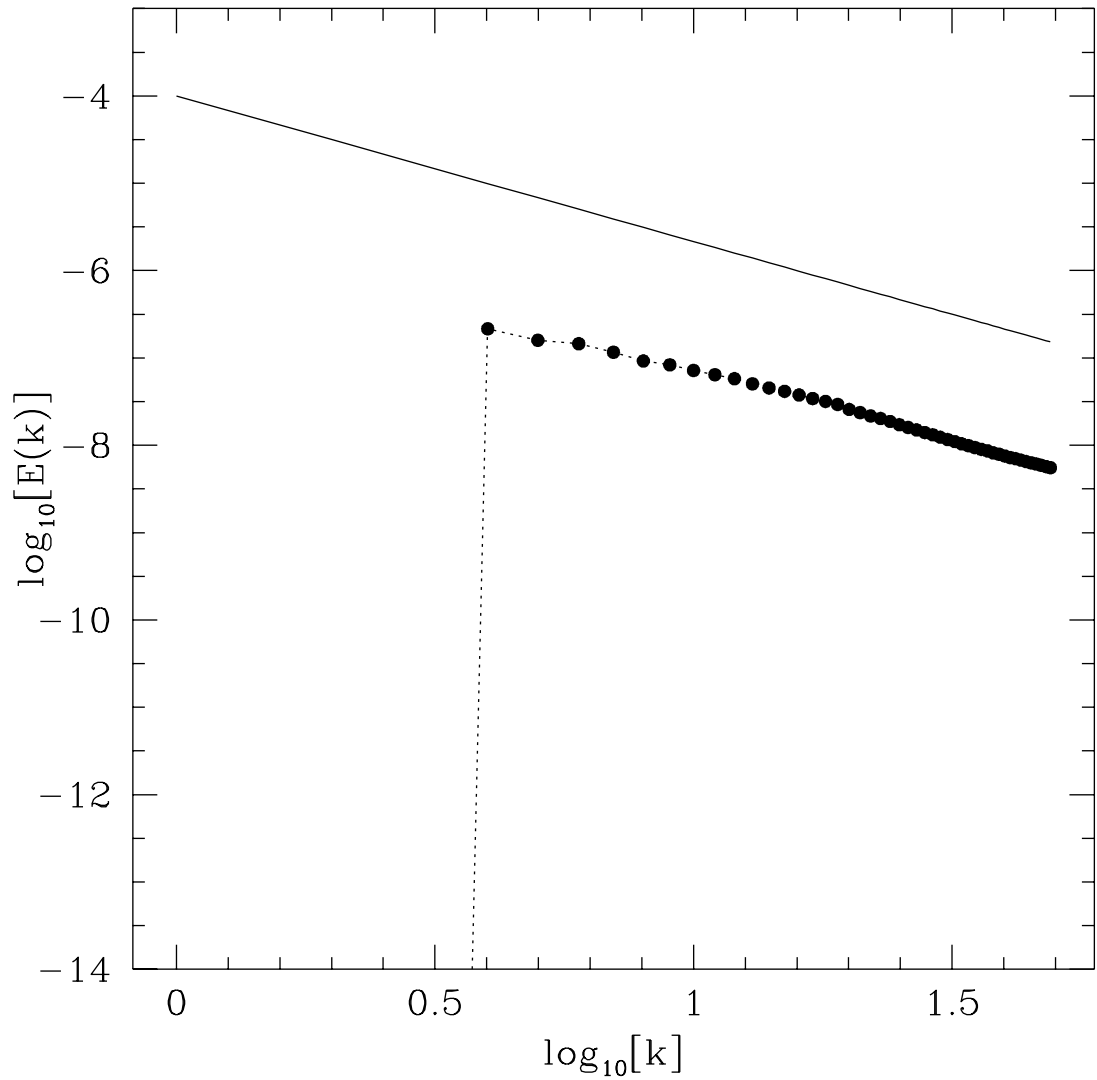,width=6.0in,height=6.0in}} 
\caption{Time averaged energy spectrum for LES of isotropic 2D
turbulence. The solid line depicts the Kolmogorov $-5/3$ slope.}
\label{fig13}
\end{figure}

\end{document}